# Modelling of Counter-Current Gas–Liquid Annular Flow in a Vertical Annulus

Mahshid Firouzi[1,2,*]

1. College of Engineering Science and Environment, The University of Newcastle, Callaghan, 2308, Australia
2. Gas & Energy Transition Research Centre, The University of Queensland, Brisbane, Queensland 4072, Australia

*E-mail: Mahshid.Firouzi@newcastle.edu.au

**Abstract**

Annular two-phase flow involving upward-moving gas and downward-flowing liquid within a vertical annular geometry is a common configuration in various engineering systems. In such systems, accurately predicting the pressure gradient is essential for understanding flow stability, equipment performance, and system design under counter-current conditions.
This paper presents a general mechanistic model to predict the pressure gradient in counter-current annular flow within a vertical annular space. The model accounts for liquid film behaviour, droplet entrainment, and interfacial shear. Model predictions show that the pressure gradient increases with rising gas flow rate. The increase in pressure gradient is attributed to a higher entrainment of liquid droplets into the gas core and enhanced interfacial forces between the phases.

*Keywords:* Counter-current, annular flow, falling film, sheared film, modelling, gas-liquid two-phase flow.

## 1 Introduction

Counter-current gas–liquid annular flow occurs when gas and liquid move in opposite directions within the same annular space between two concentric cylinders. In such systems, a liquid film typically flows downward along the inner wall of the outer cylinder, while gas flows upward through the core of the annulus. These flow regimes are encountered in a range of industrial applications including oil and gas, chemical processing, geothermal systems, and nuclear cooling, where accurate prediction of pressure profiles is critical for design and safety (Firouzi and Hashemabadi 2009, Wu et al. 2017, Amani and Firouzi 2022, Biton et al. 2022, Barabash et al. 2023).

While there is extensive literature on gas–liquid flow correlations and pressure-drop models, the majority of studies have focused on co-current flow, primarily motivated by oil and gas well applications (Hagedorn and Brown 1965, Beggs and Brill 1973, Taitel et al. 1980, Barnea et al. 1985, Abdul-Majeed and Firouzi 2019). Few studies have examined two phase gas-liquid flow through an annular geometry (Kelessidis and Dukler 1989, Caetano et al. 1992, Hasan and Kabir 1992, Lage and Time 2002, Julia and Hibiki 2011, Firouzi et al. 2016, Cao et al. 2024). Most of this work is experimental in nature, with limited development of theoretical models for flow regime transitions. For a comprehensive review of co-current vertical flows, see Wu et al. (2017).

Despite extensive literature on two-phase flow patterns, relatively few studies have addressed the hydrodynamics of counter-current flow, apart from investigations into the counter-current flow limitation (CCFL) or flooding phenomenon as exemplified by (Ma et al. 2023). One of the earliest flow maps for vertical counter-current flow was developed experimentally by Yamaguchi and Yamazaki (1982) for an air-water system. This was followed by the work of Taitel and Barnea (1983) who used a simplified mechanistic model to predict pressure gradients and flow regime transitions namely bubble, slug, and annular regimes, in vertical counter-current flow. Subsequent studies (Ghiaasiaan et al. 1997, Kim et al. 2001, Ghosh et al. 2013) continued to experimentally investigate flow patterns within counter-current pipe flows using a number of different experimental methods.

Previously, Firouzi et al. (2016) developed models to predict flow characteristics of counter-current gas-liquid flows for the bubble and slug regimes in vertical annuli, validating their modelling results with a range of experimental studies focused specifically on unconventional gas wells (Wu et al. 2019, Wu et al. 2020). While there has been some experimental work on pressure gradients and phase holdup in bubble and slug regimes for counter-current pipe flows, there is a significant absence of predictive models for counter-current annular flow in an annular geometry. In response, this paper presents a mathematical model based on a mechanistic approach to investigate the pressure gradient characteristics of the counter-current annular regime in an annular geometry.

## 2 Model description

Annular flow occurs principally at high gas flowrates and is characterised by a continuous gas phase in the core of the annulus. The liquid phase appears in forms of thin liquid films surrounding the gas core and tiny liquid droplets entrained in the gas core. Figure 1 illustrates the schematic of an annular flow regime in an annulus. The modelling approach of Ansari et al. (1990), which was validated against a wide range of well data, for co-current annular flows in pipes was employed to model a counter-current annular flow regime in an annulus.

It is assumed that the flow is at steady-state, fully developed, there is no phase change, and the entrained droplets move with the same velocity as the gas in a homogeneous mixture in the core. Tubing and casing liquid films are also assumed to have a uniform thickness.

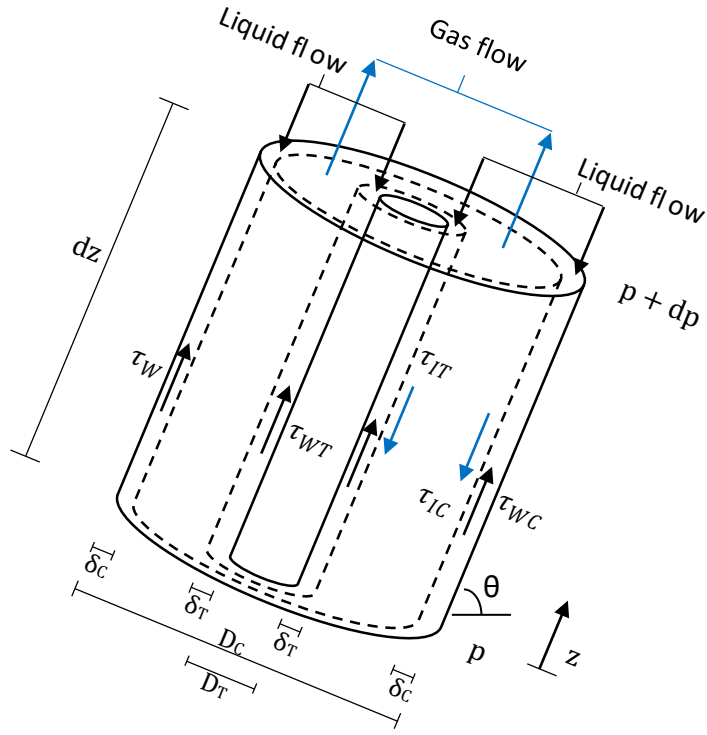

**Figure 1.** Schematic of a counter-current two-phase flow in an annulus with Θ, inclination angle between 0° and 90°

Applying momentum equations on the liquid films and the gas core results in the following equations.

$$-\frac{A_{TF}}{A_T}\left(\frac{dp}{dz}\right)_{TF} + \tau_{TW}\frac{S_{TF}}{A_T} + \tau_{TI}\frac{S_{TI}}{A_T} + \rho_L g \frac{A_{TF}}{A_T}\sin\theta = 0 \qquad (1)$$

$$-\frac{A_{CF}}{A_T}\left(\frac{dp}{dz}\right)_{CF} + \tau_{CW}\frac{S_{CF}}{A_T} + \tau_{CI}\frac{S_{CI}}{A_T} + \rho_L g \frac{A_{CF}}{A_T}\sin\theta = 0 \qquad (2)$$

$$+\frac{A_C}{A_T}\left(\frac{dp}{dz}\right)_{C} - \tau_{TI}\frac{S_{TI}}{A_T} - \tau_{CI}\frac{S_{CI}}{A_T} - \rho_c g \frac{A_C}{A_T}\sin\theta = 0 \qquad (3)$$

where $(dp/dz)_{TF}$, $(dp/dz)_{CF}$ and $(dp/dz)_C$ are the total pressure gradient of the tubing liquid film, casing liquid film and gas core. $S$ and $A$ are the perimeter and area associated with the films, gas core and gas-film interfaces. $\tau_{TW}$ and $\tau_{CW}$ are the shear stress occurring on the tubing and casing walls and $\tau_{TI}$ and $\tau_{CI}$ are the shear stress corresponding with the interface of the tubing film-core and casing film-core. The wall shear stresses are determined as follows:

$$\tau_{XF} = \rho_L f_{XF}\frac{U_{XF}^2}{8} \qquad X = T, C \qquad (1)$$

where $f_F$, the Darcy friction factor is evaluated by a Blasius type expression as in Caetano et al. (1992):

$$f_F = C \operatorname{Re}^{-x} \tag{2}$$

where *C* is a geometry coefficient and *x*=1 for laminar flows and *x*=0.25 for turbulent flows.

$U_{XF}$, the velocity of liquid in the films is evaluated from an overall liquid mass balance which gives

$$U_{CF} = \frac{U_{SL} A_T (1 - Fe - H_{TF})}{A_{CF}} \tag{3}$$

and

$$U_{TF} = \frac{U_{SL} A_T (1 - Fe - H_{CF})}{A_{TF}} \tag{4}$$

$A_T$ is the total cross sectional area of the annulus, $A_{TF}$ and $A_{CF}$ are the areas occupied by the tubing and casing films. $U_{SL}$ and $U_{SG}$ are the liquid and gas superficial velocities. $H_{TF}$ and $H_{CF}$ are the liquid holdup for the tubing film and casing film.

*Fe*, the fraction of total liquid entrained in the core, is evaluated by the correlation proposed by Cioncolini and Thome (2012) for the core Weber number $\left(\text{We} = \frac{\rho_C U_C^2 D_H}{\sigma}\right)$ greater than 10 and smaller than $10^5$ as follows:

$$Fe = \left[1 + 279.6 (\text{We})^{-0.8395}\right]^{-2.209} \tag{5}$$

Here $\rho_C$, the core flow density is a function of *Fe* as $\rho_C = \frac{\rho_L Fe\, U_{SL} + \rho_G U_{SG}}{U_{SG} + Fe\, U_{SL}}$.

The interfacial shear stress is approximated by

$$\tau_{IX} = \rho_C f_{IX} \frac{U_C^2}{8} \qquad X = T, C \tag{6}$$

Here $U_C$, the core velocity, is calculated as $U_{SC}/H_{L,core}$ in which $U_{SC}$, the core superficial velocity is defined as $U_{SC} = Fe\, U_{SL} + U_{SG}$. $f_I$, the friction factor associated with the interface is evaluated using the following expression:

$$f_{IX} = f_{CS} Z_X \qquad X = T, C \tag{7}$$

where $f_{CS}$ is the friction factor corresponding to the gas superficial velocity in the core. *Z* is a correlating factor which is defined using the following correlations suggested by Wallis (1969) for Fe>0.9 and Whalley &Hewitt (1978) for Fe<0.9.

$$Z_X = \begin{cases} (1+300\,\bar{\delta}_X) & Fe > 0.9 \\ \left(1+24\left(\dfrac{\rho_L}{\rho_G}\right)^{1/3}\bar{\delta}_X\right) & Fe < 0.9 \end{cases} \qquad X = T, C \tag{8}$$

$\bar{\delta}$ is the dimensionless film thickness. Subscripts $T$ and $C$ refer to tubing and casing. Combining equations (1)-(11) yields the following equations.

$$\dfrac{-X_M^2}{64}\left((1-Fe-H_{TF})(1+K)\right)^{2-x}\left(\dfrac{1-K}{\bar{\delta}_C(1-\bar{\delta}_C)}\right)^3 - \dfrac{(1-K)(1-K^2)^2}{a^3}\ldots$$

$$\left[Z_C\dfrac{(1-2\bar{\delta}_C)a}{4\bar{\delta}_C(1-\bar{\delta}_C)} + Z_C(1-2\bar{\delta}_C) + Z_T K(1+2\bar{\delta}_T)\right] - Y_M = 0 \tag{9}$$

$$\dfrac{-X_M^2}{64}\left((1-Fe-H_{CF})\dfrac{1+K}{K}\right)^{2-x}\left(\dfrac{1-K}{K\bar{\delta}_T(1+\bar{\delta}_T)}\right)^3 + \dfrac{(1-K)(1-K^2)^2}{a^3}\ldots$$

$$\left[Z_T\dfrac{(1+2\bar{\delta}_T)a}{4K\bar{\delta}_T(1+\bar{\delta}_T)} + Z_C(1-2\bar{\delta}_C) + Z_T K(1+2\bar{\delta}_T)\right] - Y_M = 0 \tag{10}$$

Here $K=\dfrac{D_T}{D_C}$, $\bar{\delta}_T=\dfrac{\delta_T}{D_T}$, $\bar{\delta}_C=\dfrac{\delta_C}{D_C}$, $H_{TF}=\dfrac{4\bar{\delta}_T K^2(1+\bar{\delta}_T)}{1-K^2}$, $H_{CF}=\dfrac{4\bar{\delta}_C(1-\bar{\delta}_C)}{1-K^2}$ and $a=\left[1-K^2-4\bar{\delta}_C(1-\bar{\delta}_C)-4K^2\bar{\delta}_T(1+\bar{\delta}_T)\right]$.

$X_M$ and $Y_M$, the modified Lockhart-Martinelli parameters in the above equations are defined as

$$X_M = \sqrt{\dfrac{(dp/dz)_{SL}}{(dp/dz)_{SC}}} \quad \text{and} \quad Y_M = \dfrac{(\rho_L-\rho_C)g\sin\theta}{(dp/dz)_{SC}} \quad \text{in which} \quad (dp/dz)_{SL} = \dfrac{\rho_L f_{SL} U_{SL}^2}{2D_H} \text{ and}$$

$$(dp/dz)_{SC} = \dfrac{\rho_C f_{SC} U_{SC}^2}{2D_H}.$$

Equations (12) and (13) were solved simultaneously by applying the Newton-Raphson approach to determine $\bar{\delta}_T$ and $\bar{\delta}_C$ as the unknowns of Equations (12) and (13). Once the film thicknesses are known the pressure gradient can be determined by using equations (1)-(3).

## 3 Results and discussion

Figure 2 presents the predicted pressure gradient for a counter-current annular flow for different gas and liquid superficial velocities in an annulus the same size as our experimental setup. The dashed line shows the counter-current flow limitation (flooding). The results indicate that the pressure gradient in the annular flow regime increases with the gas superficial velocity (flow rate). Figure 3 shows the predicted pressure gradient and liquid holdup for a counter-current annular flow.

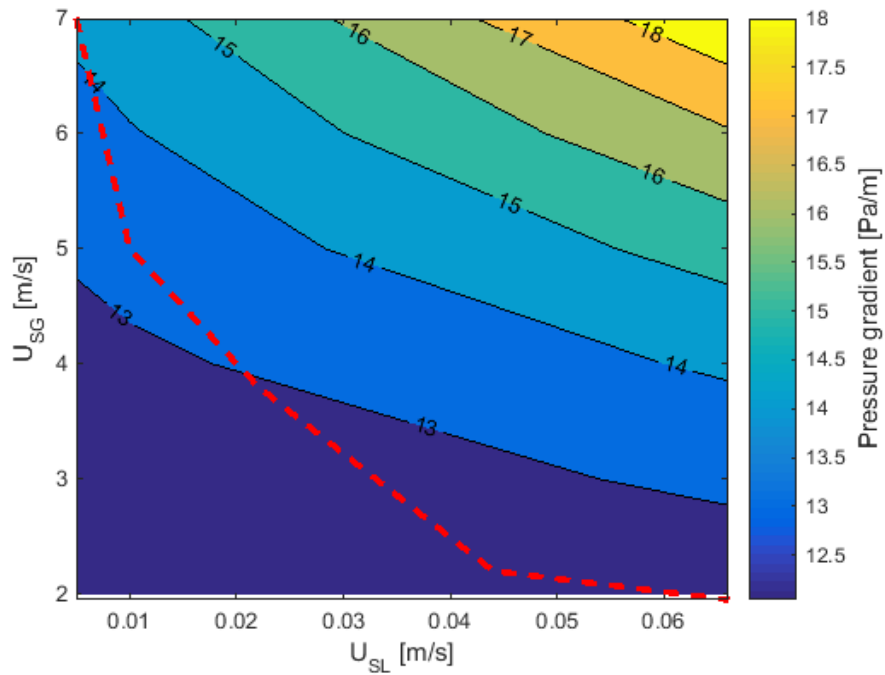

**Figure 2**. Pressure gradient of counter-current falling film flows in an annulus of $D_C$=170 mm and $D_T$=70 mm for different gas and liquid superficial velocities. The dashed line presents the counter-current flow limitation based on our experimental observations.

The predicted liquid holdup in the casing and tubing films of the annular flow remained almost constant for a constant liquid rate. However, the fraction of entrained liquid droplets in the gas core, *Fe*, increased with the gas flow rate due to the gas-liquid interfacial effect, which led to an increase in the total liquid holdup of the annular flow. Therefore, the increasing pressure gradient of annular flow regime can be partially explained through the hydrostatic (gravity) term of the total pressure gradient. The contribution of tubing/casing wall friction, tubing/casing film-core interfacial frictions as well as the contribution of gravity are shown in Figure 4(a)-(C) through equations (1)-(3). Each term was non-dimensionalised with respect to the pressure gradient of a single liquid phase of water, $\rho_L g$ . The results show an increase of the interfacial shear with increasing gas rate which is not as significant as the gravity term within the investigated range of flow rates.

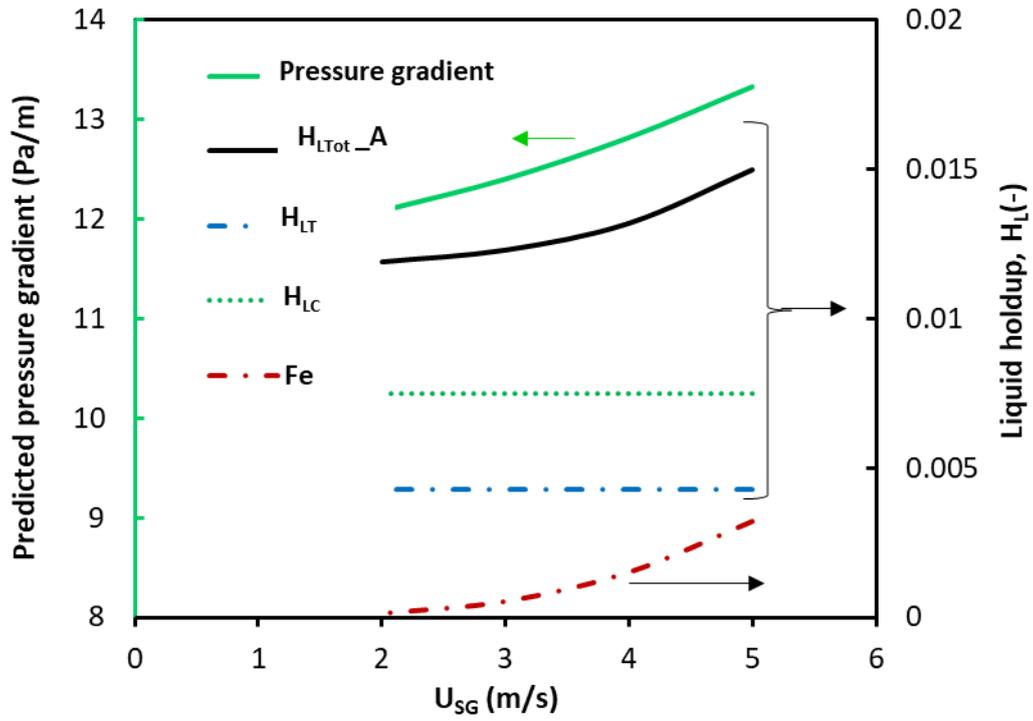

**Figure 3**. Pressure gradient and liquid holdup of counter-current annular regimes in an annulus with $D_C$=170mm and $D_T$=70mm at different gas superficial velocities and fixed liquid superficial velocity of 0.01 m/s.

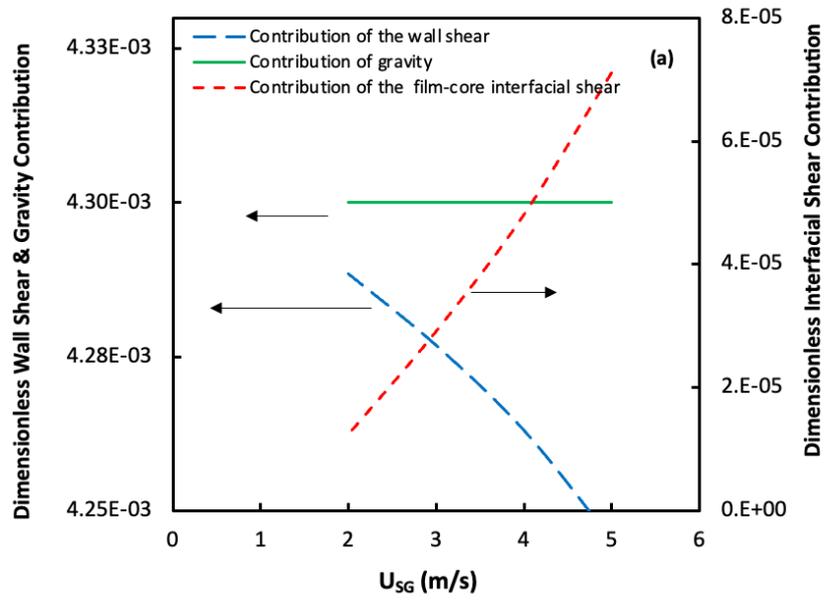
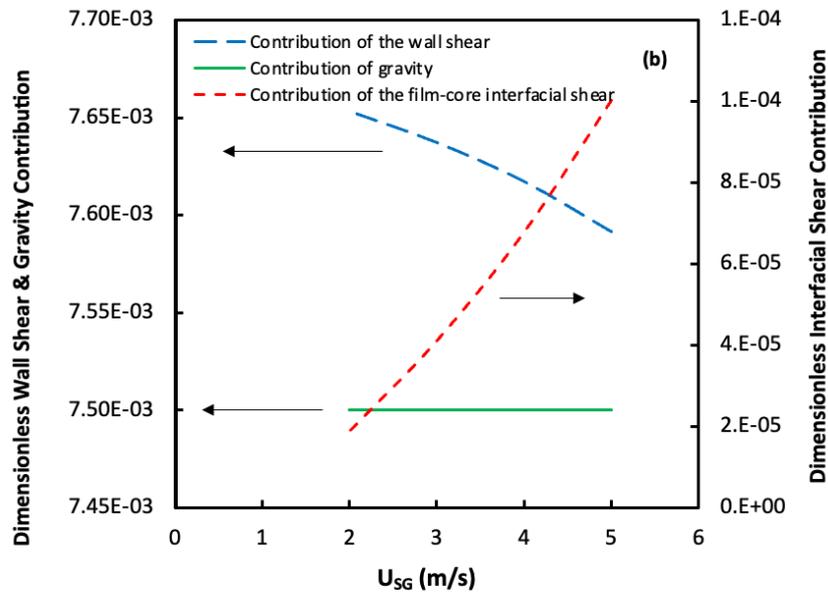
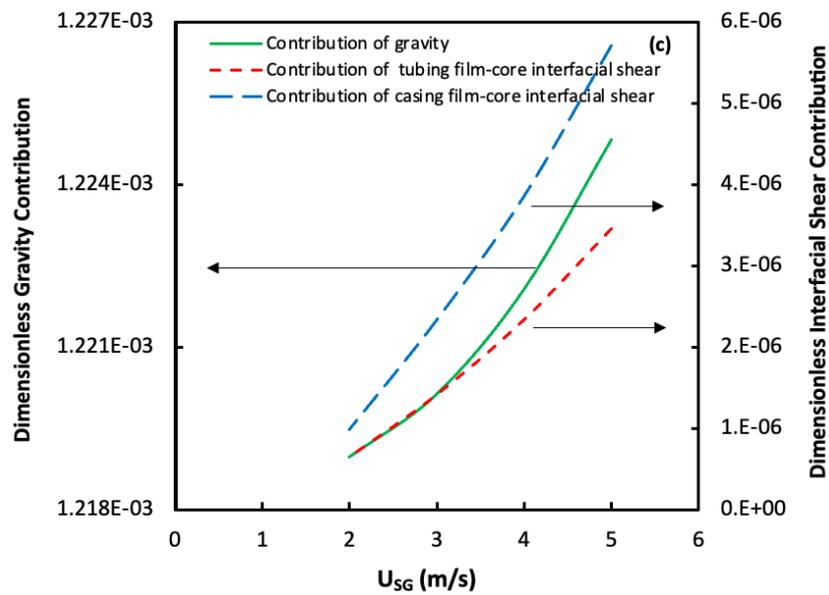

**Figure 4**. Contribution of the dimensionless shear and gravity to pressure gradient in (a) tubing, (b) casing and (C) gas core of a counter-current falling film flow in an annulus with $D_C$=170 mm and $D_T$=70 mm at a fixed superficial velocity of 0.01 m/s. All pressure gradient and shear terms are normalised by $\rho_L g$ to yield dimensionless contributions

## 4 Conclusions

This paper presented a mechanistic model to describe the flow behaviour of counter-current annular flow in an annular geometry. The model quantifies the contributions of wall shear, gas-liquid interfacial shear, and gravity term to the total pressure gradient. Results show that the fraction of entrained liquid droplets in the gas core (*Fe*) increases with gas flow rate due to enhanced interfacial shear, leading to a corresponding increase in the total liquid holdup. The results provide insights into the mechanisms driving pressure variation in counter-current annular flow and establish a basis for improved modelling of similar flow configurations in engineering applications.

**References**

Abdul-Majeed, G. H. and M. Firouzi (2019). "The suitability of the dimensionless terms used in correlating slug liquid holdup with flow parameters in viscous two-phase flows." International Communications in Heat and Mass Transfer **108**: 104323.

Amani, P. and M. Firouzi (2022). "Effect of salt and particles on the hydrodynamics of foam flows in relation to foam static characteristics." Chemical Engineering Science **254**: 117611.

Ansari, A. M., et al. (1990). Comprehensive mechanistic model for upward two-phase flow in wellbores. Proceedings - SPE Annual Technical Conference and Exhibition.

Barabash, P., et al. (2023). "Heat and mass transfer of countercurrent air-water flow in a vertical tube." Heat and Mass Transfer **59**(7): 1343-1351.

Barnea, D., et al. (1985). "Gas-liquid flow in inclined tubes: Flow pattern transitions for upward flow." Chemical Engineering Science **40**(1): 131-136.

Beggs, D. H. and J. P. Brill (1973). "A Study of Two-Phase Flow in Inclined Pipes." Journal of Petroleum Technology **25**(5): 607-617.

Biton, A., et al. (2022). "Generalized correlation for onset flooding velocity in vertical channels." International Communications in Heat and Mass Transfer **138**: 106366.

Caetano, E. F., et al. (1992). "Modeling bubble, slug, and annular flow. (Upward Vertical Two-Phase Flow through an Annulus, part 2)." Journal of Energy Resources Technology **114**(1): 14.

Cao, F., et al. (2024). "Flow regimes identification of air water counter current flow in vertical annulus using differential pressure signals and machine learning." Scientific Reports **14**(1): 12572.

Cioncolini, A. and J. R. Thome (2012). "Entrained liquid fraction prediction in adiabatic and evaporating annular two-phase flow." Nuclear Engineering and Design **243**: 200-213.

Firouzi, M. and S. H. Hashemabadi (2009). "Exact solution of two phase stratified flow through the pipes for non-Newtonian Herschel–Bulkley fluids." International Communications in Heat and Mass Transfer **36**(7): 768-775.

Firouzi, M., et al. (2016). "Developing new mechanistic models for predicting pressure gradient in coal bed methane wells." Journal of Natural Gas Science and Engineering **33**: 961-972.

Ghiaasiaan, S. M., et al. (1997). "Hydrodynamic characteristics of counter-current two-phase flow in vertical and inclined channels: effects of liquid properties." International Journal of Multiphase Flow **23**(6): 1063-1083.

Ghosh, S., et al. (2013). "Automatic classification of vertical counter-current two-phase flow by capturing hydrodynamic characteristics through objective descriptions." International Journal of Multiphase Flow **52**: 102-120.

Hagedorn, A. and K. Brown (1965). "Experimental Study of Pressure Gradients Occurring During Continuous Two-Phase Flow in Small-Diameter Vertical Conduits." Journal of Petroleum Technology **17**(4): 475-484.

Hasan, A. R. and C. S. Kabir (1992). "Two-phase flow in vertical and inclined annuli." International Journal of Multiphase Flow **18**: 279 - 293.

Julia, J. E. and T. Hibiki (2011). "Flow regime transition criteria for two-phase flow in a vertical annulus." International Journal of Heat and Fluid Flow **32**(5): 993-1004.

Kelessidis, V. C. and A. E. Dukler (1989). "Modeling flow pattern transitions for upward gas-liquid flow in vertical concentric and eccentric annuli." International Journal of Multiphase Flow **15**(2): 173-191.

Kim, H. Y., et al. (2001). "Flow pattern and flow characteristics for counter-current two-phase flow in a vertical round tube with wire-coil inserts." International Journal of Multiphase Flow **27**(12): 2063-2081.

Lage, A. and R. Time (2002). "An Experimental and Theoretical Investigation of Upward Two-Phase Flow in Annuli." SPE Journal **7**(3): 325-336.

Ma, Y., et al. (2023). "An experimental study on gas–liquid two-phase countercurrent flow limitations of vertical pipes." Experimental Thermal and Fluid Science **141**: 110789.

Taitel, Y. and D. Barnea (1983). "Counter current gas-liquid vertical flow, model for flow pattern and pressure drop." International Journal of Multiphase Flow **9**(6): 637-647.

Taitel, Y., et al. (1980). "Modelling flow pattern transitions for steady upward gas-liquid flow in vertical tubes." AIChE Journal **26**(3): 345-354.

Wallis, G. B. (1969). One-dimensional two-phase flow. New York, McGraw-Hill.

Whalley, P. B. and G. F. Hewitt (1978). The correlation of liquid entrainment fraction and entrainment rate in annular two-phase flow. Harwell, UKAEA report. **AERE-R9187**.

Wu, B., et al. (2017). "A critical review of flow maps for gas-liquid flows in vertical pipes and annuli." Chemical Engineering Journal **326**: 350-377.

Wu, B., et al. (2019). "Characteristics of counter-current gas-liquid two-phase flow and its limitations in vertical annuli." Experimental Thermal and Fluid Science **109**: 109899.

Wu, B., et al. (2020). "Use of pressure signal analysis to characterise counter-current two-phase flow regimes in annuli." Chemical Engineering Research and Design **153**: 547-561.

Yamaguchi, K. and Y. Yamazaki (1982). "Characteristics of Counter current Gas-Liquid Two-Phase Flow in Vertical Tubes." Journal of Nuclear Science and Technology **19**(12): 985-996.